\documentstyle[12pt,epsfig]{article}
\newcommand{\blankline}{\vskip .3cm}
\newcommand{\f}{\begin{equation}} \newcommand{\ff}{\end{equation}}
\begin{document} \centerline{\LARGE Matrix models as non-local
hidden variables theories} \blankline \blankline \rm
\centerline{Lee Smolin${}^*$} \blankline \centerline{\it
Perimeter Institute for Theoretical Physics} \centerline{\it
Waterloo, Ontario,  Canada N2J 2W9} \centerline{\it  \ \ and}
\centerline{\it Department of Physics, University of Waterloo}
\centerline{\it  \ \ and } \centerline{\it The Blackett
Laboratory,Imperial College of Science, Technology and Medicine }
\centerline{\it South Kensington, London SW7 2BZ, UK} \blankline
\blankline \blankline \blankline \centerline{December 15, 2001}
\blankline \blankline \blankline \blankline \centerline{ABSTRACT}
It is shown that the matrix models which give non-perturbative
definitions of string and M theory may be interpreted as
non-local hidden variables theories in which the quantum
observables are the eigenvalues of the matrices while their
entries are the non-local hidden variables. This is shown by
studying the bosonic matrix model at finite temperature, with $T$ taken
to scale as $1/N$. For large $N$ the eigenvalues of the matrices
undergo Brownian motion due to the interaction of the diagonal
elements with the off diagonal elements, giving rise to a
diffusion constant that remains finite as $N \rightarrow \infty$.
The resulting probability density and current for the eigenvalues
are then found to evolve in agreement with the Schroedinger
equation, to leading order in $1/N$, with $\hbar$ proportional to
the thermal diffusion constant for the eigenvalues.  The quantum
fluctuations and uncertainties in the eigenvalues are then
consequences of ordinary statistical fluctuations in the values
of the off-diagonal matrix elements. Furthermore, this
formulation of the quantum theory is background independent, as
the definition of the thermal ensemble makes no use of a
particular classical solution. The derivation relies on
Nelson's stochastic formulation of quantum theory, which is
expressed in terms of a variational principle. \vfill \blankline
\blankline ${}^*$ smolin@ic.ac.uk, lsmolin@perimeterinstitute.ca
\eject \tableofcontents \eject \vfill

\section{Introduction}

In this paper we describe a new proposal concerning the relationship between general
relativity and the quantum theory. This is that matrix
formulations of string, or $\cal M$
theory\cite{CH,dWHN,BFSS,IKKT}, which are known to reproduce
general relativity, at least perturbatively, are also hidden
variables theories.  More precisely, given the dynamics postulated
by these models, we show that the finite temperature classical
statistical mechanics of the matrix elements reproduces the
quantum theory of the matrix eigenvalues. This happens when the
temperature is scaled appropriately, as $T \approx 1/N$ as $N$ is
taken to infinity.  Thus, these theories contain not only a
non-perturbative definition of string theory, they contain a
reformulation of quantum theory in terms of the ordinary
statistical mechanics of a set of non-local variables which are
the matrix elements.

By a hidden variables theory we mean generally a theory which
purports to give a more detailed description of individual events
and processes for which the quantum mechanics yields only
probabilistic predictions. It is thus a theory whose ordinary
statistical mechanics reproduces, at least to a certain order of
approximation, the conventional quantum mechanics. The name hidden
variables refers to the existence of degrees of freedom beyond the
observables of the quantum theory, whose statistical fluctuations
are the source of the quantum fluctuations and uncertainties of
quantum theory.

We know from the observed violations of the Bell inequalities that
any viable hidden variable theory should be non-local.  This leads
to a natural suggestion, which is that the non-local hidden variables
describe relationships between the local degrees of
freedom\cite{meoldhidden}. A simple
hypothesis is that there is such a relational hidden variable for
each pair of degrees of freedom of a quantum theory.  This then
suggests formulating a non-local hidden variables theory in which
the fundamental, hidden degrees of freedom are elements of
matrices while the observables are the eigenvalues of the
matrices.

Such a hidden variables theory was developed some years ago and
shown to reproduce the predictions of non-relativistic quantum
mechanics for an $N$-body system\cite{meoldhidden}. When the
elements of an $N \times N$ matrix are put into a thermal bath, it
was found that under certain  conditions, the probability density
and current for the eigenvalues behaved to leading order in
$1/\sqrt{N}$, in a way that reproduced the $N$-body Schroedinger
equation.  Other non-local hidden variables have been formulated
and studied, for example, the Bohm model\cite{Bohm}.

Thus, it is not very difficult to make non-local hidden variables
theories. What is more challenging is to invent such a theory that
solves other problems besides the issues in the foundations of
quantum mechanics. Since a hidden variables theory must be
non-local, it seems very likely that a true one would have
something to do with the structure of space and time, and hence
with quantum gravity\footnote{A different proposal to unify
relativity with quantum theory in the context of a hidden
variables theory has been proposed by 't
Hooft\cite{thooft-hidden}.}.

It is then remarkable that matrix models which make use of
essentially the same idea as the hidden variables theory described
in\cite{meoldhidden}-that the real physical degrees of freedom are
matrix elements, while the eigenvalues correspond to the
observables-were developed as representations of string and $\cal
M$ theory.  These matrix formulations of string theory have solved
certain problems, but during their study a certain puzzle has
emerged concerning their quantization.

This is a consequence of the fact that one defines string theories
from compactifications of the matrix models. To define a quantum
string theory one first picks a classical solution of the matrix
model, which gives information about the background fields
including the space time geometry.  One then proceeds to define a
quantum theory with respect to that classical background.

Of course, one can define formal quantizations, for example by
means of path integral, which appear to be background independent.
However as soon as one attempts to define the quantum theory
precisely, for example, in terms of a $BRST$ quantization, one
finds that the precise definitions of the Hilbert space and $BRST$
charge $Q$ do depend on the background. Or, if one wants to define
the quantum theory precisely in a path integral context, one faces
the problem that the Lorentzian path integral does not appear to
be well defined. However, the usual solution to this problem,
which is to make a Euclidean continuation, is itself background
dependent, because it depends on a choice of time in a particular
classical spacetime manifold.  No background independent
definition of continuation between Euclidean and Lorentzian
theories is known.

This problem, of the background dependence of the quantizations, is
a major problem for string and $\cal M$ theory. As the theory
classically has many solutions, which define different spacetime
backgrounds, it seems that there ought to be a background
independent way to define the quantum theory.  If the fundamental
theory is a quantum theory it should be that the different
quantizations defined around the different backgrounds should be
approximations to a single exact quantum theory.

This is part of the motivation for the present proposal.  Perhaps
the background independent theory is not a conventional quantum
theory, but some deeper theory, which can be approximated by
conventional quantum theories when the state defines a fixed
spacetime background. Such a theory might be approached in
different ways, but here we investigate the hypothesis that it
might be a hidden variables theory. As we will show, the hint from
earlier work is correct, and a matrix model of the kind studied in
string and $\cal M$ theory can serve as a non-local hidden
variables theory which can then reproduce a quantum theory for the
eigenvalues of the matrices  to leading order in $1/N$.

That is, rather than {\it quantizing} the {\it classical} matrix
model in some conventional fashion, we will simply assume that the
off diagonal elements of the matrix model are in a classical
thermal state.  We will find that the quantum theory for the
eigenvalues can be reproduced so long as the  temperature is
scaled in a certain way with $N$.  This formulation of the quantum
theory is by definition background dependent, because the
definition of the thermal ensemble makes no reference to any
particular classical solution.

To see how this happens it is useful to consider the diagonal
elements of the matrices, which become increasingly free at low
temperatures, as analogous to the pollen grains in classical
Brownian motion.  The off diagonal elements are then analogous to
the molecules whose constant collisions with the grains cause them
to move with a Brownian motion. Indeed, the diagonal elements are
subject to random forces from their interactions with the
off-diagonal elements.  The off diagonal elements are small at low
temperature, but as we increase $N$ their effects on the diagonal
elements are greater due to their greater number. The result is
that the interaction of a diagonal elements with a large number of
off-diagonal elements introduces a Brownian motion, which is
transferred to a Brownian motion of the eigenvalues at low
temperature. That is, the randomness of the local variables-the
diagonal elements and the eigenvalues is due to their interactions
with a much larger number of non-local variables.

To find interesting behavior we have to scale $T$ in an
appropriate way with $N$ as we take the former to zero and the
latter to infinity.  In fact, we find that the model behaves
critically when we scale the temperature so that $T \approx 1/N$.
In this case the off diagonal matrix elements are of order
$1/\sqrt{N}$. However their collective effects on the diagonal
elements, and hence the eigenvalues, remain as $N \rightarrow
\infty$. One such effect is that the diffusion constant which
measures the Brownian motion of the eigenvalues remains finite as
$N \rightarrow \infty$.

Of course, the idea that quantum statistics might just be ordinary
statistics in an unusual context is an old one. In particular,
Nelson\cite{Nelson} has proposed a {\it stochastic formulation of
quantum theory} according to which the quantum description of a
particle is derived by modifying the classical description solely
by the addition of a universal Brownian motion,
which satisfies certain special properties. Chief amongst them is
that the Brownian motion is non-dissipative, in that energy and
momentum are still conserved. Nelson
shows in \cite{Nelson} that when a  classical particle is subject
to such a non-dissipative Brownian motion, its probability density
and current evolve in a way which is equivalent to that given by
the Schroedinger equation.

Nelson's formulation plays a key role in the present work, in that
we show that the stochastic formulation of quantum theory is
recovered to leading order in $1/N$, for the eigenvalues of the
matrices in our model\footnote{Note that one can quantize the free
bosonic string directly using Nelson's stochastic quantum
theory\cite{stringsnelson}.}.

In the present paper we study a bosonic matrix model, and show
that it is indeed a non-local hidden variables theory.  The
extensions of this work to the supersymmetric matrix models
associated with string and $\cal M$ theory are in progress with
Stephon Alexander\cite{superthis}.

In the next section we describe the matrix model we will study.  A
variational principle related to Nelson's formulation of quantum
theory is presented in section 3, where the basics of the theory
of Brownian motion are reviewed for those unfamiliar with it. In
section 4 we describe the basic physical picture which suggests
the connection between the classical statistical mechanics of the
matrix model and quantum theory. In section 5 we estimate the
dependence of the relevant diffusion constants on $N$, $T$ and
other parameters. Finally, the derivation of the Schroedinger
equation for the eigenvalues is given in section 6.

\section{The model}
We study a bosonic matrix model which is the bosonic part
of the models used in string and $\cal M$
theory\cite{CH,dWHN,BFSS,IKKT}.
The degrees of freedom are $d$ $N \times N$ real symmetric
matrices $X_{a i}^{\ j}$, with $a=1,...,d$ and $i,j=1,...,N$.  The
action is, \f S= { \mu} \int dt Tr \left [
\dot{X}^2_a + \omega^2 [X_a,X_b][X^a,X^b]   \right ]
\label{action} \ff We choose the matrices $X^a$ to be
dimensionless. $\omega$ is a frequency and $\mu$ has dimensions of
$\mbox{mass}\cdot \mbox{length}^2$.  We do not assume $\hbar =1$,
in fact, as we aim to derive quantum mechanics from a more
fundamental theory, $\hbar$ is not yet meaningful.  We will
introduce $\hbar$ as a function of the parameters of the theory
when we derive the Schroedinger equation as
an approximate evolution law.
We may note that the parameters of the theory define an energy
$\epsilon = \mu \omega^2$.
The basic idea is that the off diagonal matrix elements of $X^a$
will be the non-local hidden variables.  The physical observables
will be defined to be the eigenvalues $\lambda^a_i$ of the
matrices.  We will put the system at a small, but finite
temperature, the result of which will be that the matrix elements
undergo Brownian motion as they oscillate in the potential.  It
follows from linear algebra that the eigenvalues also undergo
Brownian motion.  We will see that the parameters of the theory
can be scaled with $N$ in such a way that Nelson's stochastic
formulation of quantum mechanics is realized for the eigenvalues.
When $T=0$ the matrices must commute with each other so as to
achieve the vanishing of the potential energy.
This means that it is possible to
simultaneously diagonalize them.   When $T$ is finite, but
small compared to $\epsilon$, the off diagonal elements will
on average be small. As a
result, it is useful to split the matrices into diagonal and
off-diagonal pieces, \f X_{a i}^{\ j} = D_{a i}^{\ j} + Q_{a i}^{\
j} \ff where $D^a =\mbox{diagonal}(d^a_1,...,d^a_N)$ is diagonal
and $Q_{a i}^{\ j}$ has no diagonal elements.   Since the $Q_{a i}^{\ j}$
are dimensionless we will expect them to scale like a power
of $T/\mu\omega^2$.
We then write the action \ref{action} as \f S = \int dt \left [
{\cal L}^d + {\cal L}^Q + {\cal L}^{int} \right ] \ff
The theory of the $d$'s alone is free,
\f
{\cal L}^d = { \mu}
\sum_{ai} (\dot{d}_i^a )^2 ,
\ff
while the theory of the $Q$'s
alone has the same quartic interaction \f {\cal L}^Q = { \mu}
\left [ \sum_{aij} ( \dot{Q}_{ai}^{\ j} )^2 +   \omega^2
[Q_a,Q_b][Q^a,Q^b]   \right ] \label{LQ} \ff The interaction terms
between the diagonal and off-diagonal elements are
\f {\cal
L}^{int} = {2 \mu \omega^2 } \sum_{abij} \left [ - (d^a_i -d^a_j
)^2 (Q_{bi}^{\ j})^2 - (d^a_i -d^a_j)(d^b_i -d^b_j)Q_{ai}^{\
j}Q_{bj}^{\ i} + 2 (d^a_i -d^a_j ) Q_{i}^{b j} [Q_a , Q_b ]_{j}^{\
i} \right ] \label{Lint}
\ff
We note that the model has a
translation symmetry given by \f d^a_i \rightarrow d^a_i + v^a.
\label{translations} \ff The result is that the center of mass
momentum of the system is conserved

\section{The statistical variational principle}
We now introduce the basic ideas behind the stochastic formulation
of quantum mechanics of Edward Nelson. We introduce Nelson's
formulation by means of a variational principle, which we call the
statistical variational principle. This is closely related to
variational principles previously introduced by
Guerra\cite{Guerra} and by Nelson\cite{Nelson}. We then will
formulate it in hamiltonian langauge, which will provide insight
into how the linearity of the state space of quantum theory
emerges from the theory of Brownian motion.

\subsection{$S$ ensembles}
Consider a dynamical system living on an $n$-dimensional
configuration space coordinatized by $x^a$.  The dynamics can be
described in terms of a Hamilton-Jacobi function, $S(x,t)$ which
solves the Hamilton-Jacobi equation
\f
\dot{S} + {1 \over 2m}
(\partial_a S)^2 + U =0 \label{HJ}
\ff
A particular solution $S$
defines a family of classical trajectories whose momenta at any
point $x^a$ are defined by
\f
p_a (x) =  \partial_a S
\ff
There are many solutions to the Hamilton-Jacobi equation, each
of which defines a congruence of classical trajectories.
A statistical description of the system may be given in terms of a
probability density $\rho (x,t)$ and a probability current $v^a
(x,t)$, which together satisfy
\f \dot{\rho} + \partial_a (\rho
v^a )=0
\ff
Since the probability is conserved we may always assume
$\int d^n x \rho (x,t) =1.$

Now we will do something unusual. Let us restrict attention to an
ensemble of classical trajectories whose evolution is determined
by a particular solution $S$ of the Hamilton-Jacobi equation. We
may call this an $S$-ensemble. These have a probability density
$\rho_S (x,t)$. Since the momentum is determined by $S$, so must
be the probability current. We then have, \f m v_a = \partial_a S
\label{tied} \ff so that the probability conservation equation is
now \f \dot{\rho} + {1\over m} \partial_a (\rho \partial^ a S )=0
\label{Sconserve} \ff The restriction of attention to a
statistical ensemble with fixed solution $S$ is unusual, but it
does not take us out of the domain of classical physics.
The total probability density may be recovered formally as \f
\rho_{total}= \sum_S \rho_S \ff where the sum is over all
solutions to the Hamilton-Jacobi function.  Of course, the
probability current does not add.
Now we may notice that such ensembles are given by the solutions
of a simple variational principle
\f
I[\rho, S] = \int dt \int
d^nx \rho (x,t) \left [ \dot{S} + {1\over 2m}(\partial_a S )^2 +
U(x) \right ]
\label{simple}
\ff
The equations that arise from varying $\rho$ and
$S$ are, respectively, the Hamilton-Jacobi equation (\ref{HJ}) and
the probability conservation equation (\ref{Sconserve}).
Thus, the variational principle describes an ensemble of
trajectories, each of which evolves according to the
Hamilton-Jacobi equation, so that the current velocity is
proportional to $\partial_a S $

We note that the action and equations of motion are invariant
under time reversal with
\f
   t \rightarrow -t , \ \ \  S \rightarrow -S, \ \ \  \rho
   \rightarrow \rho
   \label{timereversal}
\ff

\subsection{A brief review of the theory of Brownian motion}

Nelson's stochastic version of quantum theory may be formulated in
the language we have just introduced. To do this we assume that in
addition to the classical motion, the particles in our ensemble
are subject to a Brownian motion.  This Brownian motion is,
however, unusual, in that it does not alter the condition that
each trajectory in the ensemble is governed by the same solution
of the Hamilton-Jacobi equation.  We will see that this
requirement can be met, by altering in a small way the
Hamilton-Jacobi equation itself, to take into account the fact
that trajectories undergoing Brownian motion are not
differentiable. Nelson calls the resulting Brownian motion
dissipationless Brownian motion, as energy and momentum are still
conserved. To describe this dissipationless Brownian motion we may
use the language of stochastic differential equations\footnote{For
reviews see \cite{Nelson}.}  In this language the small change in
time of a trajectory is given by \f Dx^a = b^a (x(t),t) dt +
\Delta w^a  \ \ \  dt >0 \ff for small positive changes in time
and \f D^*x^a = -b^{*a} (x(t),t) dt + \Delta w^{*a}  \ \ \  dt <0
\ff for small negative changes in time. $b^a$ and $b^{*a}$ are
called the forward and backwards drift velocities. They describe
the average motion of the particles in the ensemble. They are
defined by \f b^a (x,t) = \lim_{\Delta t \rightarrow 0} <
{x^a(t+\Delta t) - x^a(t) \over \Delta t } >_{x(t)=x} \label{defb}
\ff and \f b^{*a} (x,t) = \lim_{\Delta t \rightarrow 0} < {x^a(t)
- x^a(t-\Delta t) \over \Delta t } >_{x(t)=x} \label{defb*} \ff
The different elements of the ensemble are distinguished by their
Brownian motion, which is given by a Markov process defined by \f
< \Delta w^a \Delta w^b >= - < \Delta w^{*a} \Delta w^{*b}> = \nu
dt q^{ab} \ff and \f < \Delta w^a \Delta w^{*b} > =0 \ff Here
$q^{ab}$ is a metric on the configuration space and $\nu$ is the
diffusion constant. The averages $<...>$ are defined with respect
to the ensemble. Thus, for any function $F(x)$ on the phase space
\f <F > = \int d^nx \rho (x,t) F(x) \ff From these basic
definitions one can derive easily the forward and backwards
Fokker-Planck equations
\f
\dot{\rho}= - \partial_a ( \rho b^a ) +
\nu \nabla^2 \rho \ff \f \dot{\rho}= - \partial_a ( \rho b^{*a} )
- \nu \nabla^2 \rho
\ff

From these the current conservation equation follows with \f v^a
= {1\over 2} ( b^a + b^{*a} ) \ff The difference between the
forward and backward drift velocities is called the osmotic
velocity. From the Fokker-Planck equation it satisfies \f u^a =
{1\over 2} (b^a - b^{*a} ) = \nu \partial^a \ln \rho \ff We thus
see that the diffusion constant measures the extent to which the
paths are non-differentiable, so that the forward and backwards
drift velocities are not equal. This is of course possible
because they are defined in eqs. (\ref{defb},\ref{defb*}) in such
a way that the limit $\Delta t \rightarrow 0$ is taken {\it
after} averaging over the ensemble.

\subsection{Quantum Brownian motion and Nelson's stochastic
formulation of quantum theory}

With this quick survey of Brownian motion over, we return to the
case of interest, which is an ensemble of trajectories that share
the same Hamilton-Jacobi function, $S$.  We want to preserve the
property (\ref{tied}) that the current velocity is proportional to
the gradient of the Hamilton-Jacobi function, but we want to find
a way to include Brownian motion within the ensemble defined by a
particular Hamilton-Jacobi function. One way to approach this is
to modify the statistical variational principle to include the
effects of Brownian motion.
It is not hard to see that this is possible, and that the right
extension of the variational principle is
\f
I^\nu [\rho, S] =
\int dt \int d^nx \rho (x,t) \left [ \dot{S} + {1\over 2m}
(\partial_a S )^2 + {m\nu^2 \over 2} (\partial_a \ln \rho )^2 +
U(x) \right ]
\label{new}
\ff
To see why let us use the fact that
$\partial_a S$ is proportional to the momentum.  Thus we have for
smooth motion, on solutions to the variational principle,
\begin{eqnarray}
\int dt \int d^nx \rho (x,t){1\over 2m}  (\partial_a S )^2 &=&
\int dt \int d^nx \rho (x,t) {1\over 2m} (p^a )^2
\nonumber \\
&=& \int dt \int d^nx \rho (x,t) {m\over 2} (\dot{x}^a )^2
\nonumber \\
&=&\int dt \int d^nx \rho (x,t) {m\over 2} \lim_{\Delta t
\rightarrow 0} \left ( {x^a(t+\Delta t) - x^a(t) \over \Delta t }
\right )^2 \end{eqnarray} However, for a Brownian motion process,
the limit in the last line is not defined. So this is not a
consistent variational principle when $\nu \neq 0$.  To define a
variational principle that is well defined for the case of
Brownian motion where the paths are non-differentiable we need to
take the limit that defines the derivative outside of the ensemble
average. Thus, we may define instead,
\f
 \int dt \int d^nx \rho (x,t) {1\over 2m} (p^a )^2
\equiv \int dt \lim_{\Delta t \rightarrow 0} \int d^nx \rho (x,t)
{m\over 2} \left ( {x^a(t+\Delta t) - x^a(t) \over \Delta t }
\right )^2 \ff

This form of the integrand appears, however, to lack invariance
under time reversals, eq. (\ref{timereversal}).
This is because when the paths are non-differentiable $b^a$ and
$b^{*a}$ may not be equal. However, we may notice that under  the
time integral we can write
\begin{eqnarray}
\int dt \lim_{\Delta t
\rightarrow 0} \int d^nx \rho (x,t) {m\over 2} \left (
{x^a(t+\Delta t) - x^a(t) \over \Delta t } \right )^2 &=& \int dt
\lim_{\Delta t \rightarrow 0} \int d^nx  {m\over 4} [\rho (x,t)
\left ( {x^a(t+\Delta t) - x^a(t) \over \Delta t } \right )^2
\nonumber \\
&&+ \rho (x,t-\Delta t) \left ( {x^a(t) - x^a(t-\Delta t) \over
\Delta t } \right )^2 ]
\end{eqnarray}
Now, we notice that \f \rho (x,t-\Delta t) = \rho (x,t) - \Delta t
\dot{\rho} (x,t)
\ff
As $\dot{\rho} (x,t)$ is given by the Fokker
Planck equation the second term leads to terms that are well
behaved and vanish as $\Delta t \rightarrow 0$. Thus, we can take
the average and then the limit, using (\ref{defb},\ref{defb*}) to
find,
\begin{eqnarray}
\int dt \lim_{\Delta t \rightarrow 0} \int d^nx \rho (x,t) {m\over
2} \left ( {x^a(t+\Delta t) - x^a(t) \over \Delta t } \right )^2
&=& \int dt \int d^nx \rho (x,t) {m\over 4} [(b^a )^2 + (b^{*a})^2
+C ]
\nonumber \\
&=& \int dt \int d^nx \rho (x,t) {m\over 2} [(v^a )^2 + (u^{a})^2
+ C ]   \nonumber \\
\end{eqnarray} Here $C$ is an infinite constant, which is equal to
\f C = { \nu d }   \lim_{\Delta t \rightarrow 0} {1 \over \Delta
t} \label{C} \ff Thus, we have found that we can extract an
infinite constant from the action, leaving us with a finite piece
that is well defined and time reversal invariant.  So we have, \f
\int dt \lim_{\Delta t \rightarrow 0} \int d^nx \rho (x,t)
{m\over 2} \left ( {x^a(t+\Delta t) - x^a(t) \over \Delta t }
\right )^2 = \int dt \int d^nx \rho (x,t) [{1\over 2m}
(\partial_a S )^2 + {m\nu^2 \over 2} (\partial_a \ln \rho )^2]  +
{ m C \over 2} \ff Constants, even infinite constants, play no
role in classical action principles. Hence, putting this last
result back in the definition of the action principle we see that
the effect of exchanging the order of the integral and the limit
is the new action principle (\ref{new}). Thus we see that while
we can have a Brownian motion within the trajectories defined by
a given Hamilton-Jacobi functional, the definition of Brownian
motion requires that we modify the Hamilton-Jacobi equation, so
that we have a consistent variational principle, even in the
presence of non-differentiable paths.

From the argument just given we see that the variational
principle (\ref{new}) is  equivalent to defining the time
derivatives so that they are taken outside the ensemble
averages.  This yields the standard variational principle when
the trajectories are smooth, because then it doesn't matter in
which order we take the limits involved in the definition of the
time derivatives and the ensemble average. But for
non-differentiable paths the order matters and we see that we
must take the limit defining the time derivative after that
defining the ensemble average. As we have seen, up to an infinite
constant which may be ignored, this is equivalent to adding the
new term ${m\nu^2 \over 2} (\partial_a \ln \rho )^2 $ to the
variational principle.

The new Hamilton-Jacobi equation follows from (\ref{new}) by
varying by $\rho$. We find that
\f
\dot{S} + {1 \over 2m}
(\partial_a S)^2 - 2m\nu^2 {1\over \sqrt{\rho}} \nabla^2
\sqrt{\rho} + U =0
\label{newHJ}
\ff
The Hamilton-Jacobi
functional is thus modified by a new potential term which is a
function of the probability density. We recall that this unusual
feature follows because it is the gradient of the probability
density that measures the importance of the non-differentiability
of the paths.
The probability conservation, however is not modified.
Now, the big surprise is that (\ref{newHJ}) and (\ref{Sconserve})
are nothing else than the real and imaginary parts of the
Schroedinger equation, with
\f
\Psi (x,t) = \sqrt{\rho (x,t)}
e^{\imath S /\hbar}
\label{Psi}
\ff
and with
\f
\hbar = 2\nu m.
\label{hbar}
\ff
Thus, the
variational principle (\ref{new}), in the presence of Brownian
motion is equivalent to \f \imath \hbar {d \Psi (x,t) \over dt} =
\left [ - {\hbar^2 \over 2m} \nabla^2 + U(x) \right ]\Psi (x,t)
\ff One way to understand this is the following. From the point of
view presented here, a quantum state is nothing more nor less than
an ordinary statistical ensemble of Brownian motion trajectories,
which share a single hamilton-jacobi function, $S$, where that $S$
is itself a solution to a modified Hamilton-Jacobi equation,
modified to take into account the change in the definition of the
momentum necessary when the motion is Brownian. This is the basic
message of Nelson's stochastic quantum theory.

\subsection{Hamiltonian formulation of the statistical variational principle}
To find the hamiltonian formulation of the statistical variational
principle note that (\ref{new}) can be written
\f
 I^\nu [\rho, S] = \int dt \int d^nx \left [ S \dot{\rho} -
{\cal H}[\rho,S,x]  \right ] \label{hamform} \ff where the
Hamiltonian density is \f {\cal H}[\rho,S,x] = \rho    \left [
{1\over 2m}  (\partial_a S )^2 + {m\nu^2 \over 2} (\partial_a \ln
\rho )^2 + U(x) \right ]. \label{hamiltonian}
 \ff
Thus we see that the probability density $\rho$  can be considered as
a conjugate coordinate with  $S$ its conjugate momentum, so that we have
an infinite dimensional phase space with
\f
\{ \rho (x) , S(x^\prime ) \} = \delta^n (x^\prime , x )
\ff
The Hamiltonian,
\f
H= \int d^nx {\cal H}
\ff
is then conserved in time.   It is easy to check that Hamilton's equations of motion
are the Hamilton-Jacobi equation, (\ref{newHJ}) and the
probability conservation equation (\ref{Sconserve}).  We note that
this is true for
any value of the diffusion constant $\nu$ so that this is
true in both classical and
quantum theory.
To get more insight into how the linearity of quantum theory has
emerged from the theory of Brownian motion, we can write out the
conserved hamiltonian, \f H= \int d^nx \rho \left [ {1 \over 2m}
(\partial_a S )^2 + {m\nu^2 \over 2} (\partial_a \ln \rho )^2 +
U(x) \right ] \ff This seems a very non-linear expression, but it
is nothing but the expectation value of a linear operator. To see
this we rewrite it slightly as
\f
H= \int d^nx \sqrt{\rho}
e^{-\imath S/\hbar} \left [  {1 \over 2m}(\partial_a S )^2 - {2
m\nu^2 } { \nabla^2 \sqrt{\rho} \over \sqrt{\rho}} + U(x) \right ]
\sqrt{\rho} e^{\imath S/\hbar}
\ff
Using (\ref{Psi}) and
(\ref{hbar}) this is easily seen to be equal to
\f
H = \int d^n x
\bar{\Psi } \hat{\cal H} \Psi \ff with \f \hat{\cal H} = -{\hbar^2
\over 2m} \nabla^2 + U(x).
\ff
Thus, the very non-linear seeming
equation $\dot{H}=0$ is seen to be actually equivalent to the
linear Schroedinger equation.  We further see that the conserved
Hamiltonian which arises from the statistical variational
principle is exactly equal  to the expectation value of the
Hamiltonian operator in the quantum theory.  Thus, the
conservation of the Hamiltonian in the statistical variational
principle is equivalent to the conservation of the expectation
value of the Hamiltonian operator in the quantum theory.
\section{The physical picture}
We will now return to the matrix model we introduced in section 2.
We will show
that the ordinary statistical physics of this model has a critical
behavior when
the off diagonal sector is heated to finite temperature and the large
$N$ limit is
taken with the temperature scaled so that $T \approx 1/N$.  We will further
see that
a feature of the critical behavior in this limit is to reproduce quantum
mechanics.
That is, to leading order in $1/N$ the evolution of the probability density
and current
for the eigenvalues of the matrixes is equivalent to that given by the free
Schroedinger equation.
To show this we will apply what we have learned in the last section to the matrix model.
We will formulation an $S$ ensemble for the matrix model in terms of the statistical
variational principle.  At the fundamental level the dynamics is formulated in terms
of the variational principle without Brownian motion, i.e. the simple variational principle
(\ref{simple}).   Thus, the whole matrix model is in an ordinary statistical ensemble.
In the next section we will study the behavior the matrix model at
low temperature and large $N$. We see that when we pick $T \approx
1/N$ the off diagonal elements scale as $1\sqrt{N}$; this makes
sense as they must go to zero at $T=0$.  We also see that to
leading order the off diagonal elements can be seen to move
harmonically in an average field given by the average values of
all the other off-diagonal elements.  The diagonal elements are
not required to vanish as $T \rightarrow 0$, so they remain of
order unity. However, the diagonal elements move in a random
potential given by the oscillations of all the off diagonal
elements. The result is that the diagonal elements pick up a
random Brownian motion, on top of their free motion.   This
Brownian motion is then also experienced by the eigenvalues. We
will see that when the model is scaled critically, the diffusion
constants for the diagonal elements and eigenvalues go to constant
limits as $N \rightarrow \infty$ and $T \rightarrow 0$.
We then want to study the effect of the Brownian motion of the eigenvalues.  To do this
we derive an effective statistical action for the probability distribution of the
eigenvalues by averaging the statistical variational principle for the whole model
over the values of the matrix elements.  This is the task of section 6.  We see that the
Brownian motion term
in (\ref{new}) emerges naturally as a term in the effective statistical action for the
eigenvalues, as a result of the induced Brownian motion just described.  Furthermore, in
that limit the conserved energy of the variational principle of the whole model reduces
to the conservation of the expectation value of the free hamiltonian operator for the
eigenvalues.  Thus, in the large
$N$ limit Nelson's stochastic formulation of quantum theory emerges naturally as a
description of the statistical behavior of the eigenvalues.
\section{Estimates for the diffusion constants at low temperature}
In this section we thus investigate the consequences of putting
the matrix model in a thermal bath at a temperature $T$. We have
two tasks. The first is to understand how various quantities of
interest  scale with $T$ and $N$, at low temperatures. The second
is to derive estimates for the diffusion constants for the matrix
elements and eigenvalues that are good for low temperature and
large $N$.
By low temperature, $T$, we will mean that the ratio $ T /\mu
\omega^2$ is small.  It will be convenient to scale this ratio
with $N$, so we define \f {T \over 8 (d-1) \mu \omega^2} = {
t\over N^p} \ff with $t$ dimensionless and $p$ a power.  The
factor of $8 (d-1)$ is inserted for later convenience.
We begin by recalling how the potential is written in terms of
diagonal and off
diagonal elements (eq. (\ref{Lint})).
\begin{eqnarray}
U(d,Q)&=& {\cal L}_{int} = \mu \omega^2 Tr \left [
-2(d_a^i-d_a^j)(d^{ai}-d^{aj})Q_{bij}Q^{bij}
+2(d_a^i-d_a^j)(d^{bi}-d^{bj})Q_{bij}Q^{aji} \right.
\nonumber\\
&&+ \left. 4(d_a^i-d_a^j)Q_{bij}[Q^a,Q^b]^{ji} +[Q^a,Q^b]^{2}
\right ]
\end{eqnarray}
The classical equations of motion are then, \f \ddot{d}_a^i =
\omega^2 \left [ -8 (d_a^i-d_a^j) Q_{bij}Q^{bij}  +8
(d^{bi}-d^{bj})Q_{bij}Q^{aji} +8 Q_{bij}[Q^a,Q^b]^{ji} +4 [Q_b
,[Q^a,Q^b]_{ij} \right ] \label{dforce} \ff \f
\ddot{Q}^a_{ij}=\omega^2 \left [ -4 (\delta^a_b (d_a^i-d_a^j)^2 -
(d_a^i-d_a^j)(d^{bi}-d^{bj})) {Q}^b_{ij} +4 (d_a^i-d_a^j)Q_{bij}^2
-8 (d_b^i-d_b^j)[Q_a,Q_b]_{ij} \right ] \label{qforce} \ff
Now we will consider how each matrix element moves in an effective
potential created by the averaged motions of the other elements.
To see this we make a mean field approximation, good at large
$N$. We assume that the statistical averages
satisfy relations consistent with the symmetry of the theory.
This gives us
\f
<Q_a^{ij} Q_b^{kl} > = q^2 \delta_{ab}
(\delta_{ik}\delta_{jl}+ \delta_{il}\delta_{jk} )
\ff
\f
<(d_a^{i} -d_a^j)(d_b^{k} -d_b^l)> = r^2 \delta_{ab}
(\delta^{ik}\delta^{jl} -\delta^{il}\delta^{jk})
\ff
\f
<Q_{aij}>=<d_a^i> =
<dQ>=<Q^3> = <d^3>=0
\ff
We assume also that averages of four
matrix elements factor into pairs of averages of two in all ways.
Our goal will be to solve for the value of $q$ when the off
diagonal elements are in a thermal bath.
We now write out the effective potential for the matrix elements
moving in the averaged fields of the other elements, to quadratic
order. This gives us,
\f
<U> =  {\mu \Omega_Q^2 \over 2}
Q_{aij}Q^{aij} + {\mu \Omega_d^2 \over 2} ( d_{a}^i -d_a^j )^2
\ff
with
\f
\Omega_Q^2 = 4(d-1)\omega^2 [(N-1)q^2 + 2r^2 ] \ff \f
\Omega_d^2 = 4(d-1) \omega^2 q^2 \ff Thus, each off diagonal
element moves in a harmonic potential created by the averaged
values of the other elements. The diagonal elements are, to
leading order a system of points, each connected to all the others
by springs with the same spring constant.
We will make no assumption about the statistical distribution of
the diagonal elements.  We will see that this is consistent so
long as $p$ is chosen so that the diagonal spring constant
$\Omega_d$ vanishes as $N \rightarrow \infty$.
We will assume that the $Q_{aij}$'s are in thermal equilibrium
with each other. This implies that at temperature $T$, \f {\mu
\over 2} q^2 \Omega^2_Q = {T \over 4} \ff which tells us that \f
{T \over 8 (d-1) \mu \omega^2 } = (N-1) q^4 + 2r^2 q^2 \ff
Now let us assume that we can neglect the term in $r^2q^2$ for
estimating the order of magnitudes of quantities and the
dependence on $N$. This will be consistent so long as $r^2 \leq N
q^2$.  We then arrive at the estimate that for large $N$,
\f
q = {1 \over N^{1/4}} \left ({T \over 8 (d-1) \mu \omega^2 }  \right
)^{1/4} = t^{1/4} {1 \over N^{p+1 \over 4}}  .
\label{qest}
\ff
We see that for
$p=1$, it is true that $N q^2$ is of order unity, so this is
consistent also with $r$ being of order unity as $N \rightarrow
\infty$ and $T \rightarrow 0$.
Note that we now have
\f
\Omega_Q^2 = 4 \omega^2 (d-1) [2r^2 +
\sqrt{t} N^{1-p\over 2} ]
\ff
We see that for $\Omega_Q$ not to
diverge as $N\rightarrow \infty$ we must have $p \geq 1$.  We then
see that $p=1$ corresponds to a critical point at which $\Omega_Q$
is fixed as $N \rightarrow \infty$.  We also see
from (\ref{qest}) that this corresponds
to $q \approx 1/\sqrt{N}$.
We see also that, \f {\Omega^2_d \over \Omega_Q^2 } = {q^2 \over
Nq^2 +2r^2} = {1 \over N +2{r^2 \over q^2}}
\label{Omegad}
\ff
Thus we see that if we choose $p=1$ so $\Omega_Q$ is fixed in the
limit $N \rightarrow \infty$ then in the same limit $\Omega_d$
vanishes, so the diagonal elements remain free in the mean field
approximation. From now on, unless otherwise specified, we will
take $p=1$.
Now we would like to estimate the effects on the matrix elements
of the fluctuations around the thermal averages for small
temperature and large $N$. Let us consider for example the
contribution of the last term in eq. (\ref{dforce}) to the motion
of $d_a^i$.   We have, neglecting the other terms,
\f
\ddot{d}_a^i =4  \omega^2 \sum_{bj} [Q_b ,[Q^a,Q^b]]_{ij}
\ff
Counting the
commutators, this is a sum of $dN^3$ terms, for each $a$ and $i$.
These add with random signs.  Each impulse of a fixed sign lasts
an average time of $\Omega_Q^{-1}$, as this is the time over which
the values of the $Q_{aij}$'s oscillate. The result is that
$d_a^i$ has a random Brownian motion on top of the free motion
given by the $N \rightarrow \infty$ limit.
To estimate the diffusion constant that results note that if, in a
time $\Delta t$ the average displacement resulting from the random
forces is $\Delta d$, the diffusion constant is
\f
\nu_d =
{(\Delta d )^2 \over \Delta t}
\ff
If $a_{total}$ is the total
acceleration given by the sum of the random forces over the time
$\Delta t$ we have  $\Delta d = {1 \over 2} a_{total} (\Delta
t)^2$. Thus, we have, \f \nu_d = {1 \over 4} a_{total}^2 (\Delta
t)^3
\ff
For $ a_{total}$ we may take
\f
a_{total} = \omega^2 q^3
\sqrt{N^3 d}
\ff
because we are adding $N^3 d$ terms with random
signs and average magnitude $\omega^2 q^3$.  Taking $\Delta t =
\Omega_Q^{-1}$ we have for large $N$,
\f
\nu_d = { \omega d \over
4(d-1)^{3/2}} {q^3N^{3/2} \over [1+ {2r^2 \over Nq^2 }]^{3/2} } =
{ \omega d t^{3/4} \over 4(d-1)^{3/2} N^{{3\over 4}(p-1)}  } {1
\over [1+ {2r^2 \over Nq^2 }]^{3/2} }
\ff
So we see that if we
pick $p=1$, and so long as $2r^2$ is of order one, the diffusion
constant for the diagonal elements goes to a limit which is
$N$-independent and hence $T$ independent as $N \rightarrow
\infty$ and $T \rightarrow 0$.
Under these same conditions we can show that as $N\rightarrow
\infty$ these terms make the dominant contribution to the random
forces on the diagonal elements coming from eq. (\ref{dforce}).  A
similar analysis starting from eq. (\ref{qforce}) allows us to
estimate the diffusion constant for the $Q$'s coming from the
random forces to be,
\f
\nu_Q = \omega { r^2 t \over N^{7p/4 + 5/4}  }
\ff
The result
is that for $p \geq 1$, $\nu_Q$ vanishes in the limit $N
\rightarrow \infty$.
Now a classic result of random matrix theory is that if the matrix
elements of a matrix undergo Brownian motion, so do its
eigenvalues. In a case like ours it is clear, as the eigenvalues
will be close to the diagonal values, as the off diagonal values
vanish as $N \rightarrow \infty$. But we have to be careful about
the contributions from higher order terms in perturbation theory.
By making use of the standard formula,
\f
\lambda^a_i  = d^a_I +
\sum_j {Q^a_{ij}Q^a_{ji} \over d^a_i -d^a_j } +...
\ff
we can show that the diffusion constant for the eigenvalues is
given by
\f
\nu_\lambda = \nu_d + {N \nu_Q q^2 \over r^2 }
+... \approx \omega \left [ {d t^{3/2} \over 4(d-1)^{3/2} N^{{3
\over 4} (p-1)}} + {t^{3/2} \over N^{9p/4 + 3/4}} \right ]
\label{nulambda}
\ff
This tends to $\nu_d$ as $N \rightarrow \infty$. Thus, we see that
the dominant contribution to the Brownian motion of the
eigenvalues comes from the Brownian motion of the diagonal
elements. These in turn are fluctuating because they are perturbed
by their interactions with the off diagonal elements, which are
moving in a harmonic potential, created by their averaged values
at finite temperature.   The result is that a randomness is
introduced into the motions of the eigenvalues, coming from the
interactions of the diagonal elements with a very large number of
random variables, which are the off diagonal elements. This then
illustrates that idea that  a local degree of freedom can have its motion
randomized by interaction with a large number of non-local degrees
of freedom. In the next section we will see that this may result in
behavior that for large $N$ is indistinguishable from that
predicted by the Schroedinger equation.
\section{Derivation of the Schroedinger equation}
We are now ready to derive the Schroedinger equation for the
eigenvalues of the matrices.  As we described above this is a three
step process,
\begin{itemize}
\item{}{\bf STEP 1} Formulate the statistical variational principle for the matrix model.
\item{}{\bf STEP 2} Make assumptions about the statistical ensemble. In particular, we
assume that the model is in an $S$-ensemble, heat it to finite temperature $T$ and then
study the large $N$ limit with $T \approx 1/N$.
\item{}{\bf STEP 3} Derive an effective statistical variational principle for the
eigenvalues by averaging over the variational principle of the matrix elements and
show that when $N \rightarrow \infty$ this is equivalent to Schroedinger quantum
theory for the eigenvalues.
\end{itemize}
\subsection{STEP 1:  The statistical variational principle for the matrix model}
We begin by defining an $S$-ensemble for the matrix elements. That
is, we begin with the variational principle \f I[\rho, S] = \int
dt \int (dd) (dQ) \rho (d,Q,t) \left [ \dot{S}(d,Q) + {1 \over
2\mu}  ({\delta S(d,Q,t) \over \delta d_i^a} )^2 + {1 \over
2\mu}({\delta S(d,Q,t) \over \delta Q_{ij}^a} )^2 + U(d,Q) \right
] \label{mp} \ff where $U(d,Q)$ is the interaction term ${\cal
L}^{int}$ given by (\ref{Lint}).
\subsection{STEP 2: Physical assumptions}
We than will state the physical assumptions we make concerning $\rho$ and
$S$. These are assumed only to hold to leading order in $1/N$
\begin{itemize}
\item{}The $Q$ system is in a distribution that is to leading
order in $1/N$ statistically independent of the distribution of
the eigenvalues. This means that to leading order the probability
density factorizes \f \rho (d,Q)= \rho_d (d) \rho_Q (Q) + O(1/N)
\label{factorrho} \ff
\item{}The $Q$ subsystem is in thermal equilibrium at a
temperature $T$.  So we have
\f
\rho_Q (Q) ={1 \over Z}
e^{-H(Q)/T} \label{thermal} \ff where $H(Q)$ is the hamiltonian
corresponding to the $Q$ system alone \f H(Q)= { \mu}  \left [
\sum_{aij} ( \dot{Q}_{ai}^{\ j} )^2 -  \omega^2 [Q_a,Q_b][Q^a,Q^b]
\right ]
\ff
and
\f
Z= \int dQ e^{-H(Q)/T}
\ff
\end{itemize}
As a result of these assumptions our variational principle reads,
 \f
I[\rho_d, S,T] = \int dt \int(dd) (dQ)
\rho_d (d) \rho_Q (Q)
\left [ \dot{S} (d,Q) + {1 \over 2\mu} ({\delta S \over \delta
d_i^a} )^2   + {1 \over 2\mu}({\delta S \over \delta Q_{ij}^a} )^2 + U(d,Q) \right ]
\label{Tprinciple}
\ff
\subsection{STEP 3: Derive an effective variational principle for the eigenvalues}
Now we want to derive an effective variational principle to describe the evolution
of the probability distribution for the eigenvalues. We will do this by averaging
the variational principle (\ref{Tprinciple}) over the values of the matrix elements,
and then extracting the leading behavior for large $N$ and small $T$.
We begin by inserting  the factor unity in the form
\f
1= \int
\prod_{ai} d\lambda^a_i  \delta \left (\lambda^a_i -d^a_i - \sum_j
{Q^a_{ij}Q^a_{ji} \over d^a_i -d^a_j } + ... \right )
\ff
Thus, we have,
\begin{eqnarray}
I[\rho_d, S,T] &=& {1 \over Z} \int dt \int dd dQ \int d\lambda \delta
\left (\lambda^a_i -d^a_i - \sum_j {Q^a_{ij}Q^a_{ji} \over
\lambda^a_i -\lambda^a_j } + ... \right )
\rho_d e^{-H(Q)/T}  \nonumber \\
&& \left [ \dot{S}  +  {1 \over 2\mu}({\delta S \over \delta d_i^a} )^2 +
+  {1 \over 2\mu}({\delta S\over \delta Q_{ij}^a} )^2 +
U(d,Q) \right ]
\end{eqnarray}
Now we would like to integrate over the $d$'s, which will express
the action in terms of only the $\lambda$'s and $Q$'s.  However,
before doing this we need to take into account that while the
to the order we are working, the $d$'s and the
$\lambda$'s will be undergoing Brownian motion because the
diagonal elements are moving in a random potential given by the
values of the $Q$'s. There are also additional contributions to the
diffusion constant coming from the terms in $Q$ that contribute to the eigenvalues
at higher order . So we will have to be careful about the definitions of the
velocities. In particular, we
will have to recall that in the theory of stochastic processes the
limits which define time derivatives are taken {\it after} the
averages over probability distributions, not before. So as before we must
write
\begin{eqnarray}
\int dt & \int & dd dQ \rho_d (d) \rho_Q (Q) {1 \over 2\mu}
({\delta S \over \delta d_i^a} )^2
= \int dt dd dQ \rho_d (d) \rho_Q (Q)) \mu ( V(d)^a_i )^2 \nonumber \\
&=&  \int dt  \lim_{\Delta t \rightarrow 0} \int dd dQ \rho_d (d) \rho_Q (Q) \mu \left
( {d^a_i (t + \Delta t ) - d^a_i (t) )^2  \over \Delta t^2 }
\right )
 \\
&=&  \int dt  \lim_{\Delta t \rightarrow 0} \int dd dQ \rho_d (d) \rho_Q (Q)
{\mu \over 2} \left \{ \left ( {d^a_i (t + \Delta t ) - d^a_i (t) )^2
\over \Delta t^2 } \right ) + \left ( {d^a_i (t ) - d^a_i
(t-\Delta t ) )^2  \over \Delta t^2 } \right ) \right \} \nonumber
\end{eqnarray}
Note that the last equation follows trivially, for smooth motion,
but it will have non trivial consequences once we have averaged over
the $Q$'s because the result for large $N$ is to induce Brownian motion
for the off diagonal elements and eigenvalues.
Now we perform the integral over the $d$'s. It is useful to
write
\f
d^a_i = \lambda^a_i + \Delta \lambda^a_i \ff where \f
\Delta \lambda^a_i (Q,\lambda ) = - \sum_j {Q^a_{ij}Q^a_{ji} \over
\lambda^a_i -\lambda^a_j } +...
\ff
has to be treated as a
stochastic variable, taking into account its dependence on the
$Q$'s which are themselves fluctuating due to the assumption that
they are in equilibrium in a potential.
We then have, to leading order in $1/N$,
\begin{eqnarray}
I[\rho_d, S_d,T] &=& \int dt \int  d\lambda
\rho_d (\lambda, t) \int dQ \rho_Q (Q)  \nonumber \\
&& \left [ \dot{S} (\lambda + \Delta \lambda , Q)
+ {1 \over 2\mu}({\delta S(\lambda + \Delta \lambda ,Q) \over \delta Q_{ij}^a} )^2
+ U(\lambda ,Q)
\right ]   + K.E.
\end{eqnarray}
where to leading order, the kinetic energy terms for the $d$'s
have become,
\f
K.E.= \int dt  \lim_{\Delta t \rightarrow 0}  \int
d\lambda \rho_d (\lambda, t) \int dQ \rho_Q (Q) {\mu \over 2}
\left \{ \left ( {( \lambda^a_i (t + \Delta t ) - \lambda^a_i (t)
)^2  \over \Delta t^2 } \right ) + \left ( { ( \lambda^a_i (t ) -
\lambda^a_i (t-\Delta t ) )^2  \over \Delta t^2 } \right ) \right
\}
\ff
We now are ready to integrate over the $Q$'s.  The key point is
that the dependence of the $\lambda$'s on the $Q$'s through a sum
of a large number of independent terms, $\sum_j {Q^a_{ij}Q^a_{ji}
\over d^a_i -d^a_j }$, as well as the coupling of the $\lambda$'s with the
$Q$'s coming from the terms in $U(\lambda, Q)$  turns the $\lambda$'s into
stochastic variables, described by a stochastic differential
equation of the form
\f
D\lambda^a_i = b^a_i (\lambda , t ) dt +
\Delta \lambda^a_i  \ \ \ \  \Delta t >0 \ff \f D\lambda^a_i =
b^{*a}_i (\lambda , t ) dt + \Delta^* \lambda^a_i  \ \ \ \  \Delta
t <0 \ff with \f < \Delta \lambda^a_i \Delta \lambda^b_j  > =
\delta^{ab}\delta_{ij} \nu_\lambda dt   \\\\ dt>0 \ff \f <
\Delta^* \lambda^a_i \Delta^* \lambda^b_j  > = -
\delta^{ab}\delta_{ij} \nu_\lambda dt  \ \ \ \  dt<0 \ff Here the
brackets mean \f < F(\lambda, Q) > = \int dQ \rho_Q (Q) F(\lambda
, Q) \ff
We note that we can use the value of $\nu_\lambda$, (\ref{nulambda}) computed in the
last section as all the assumptions we made there have been carried over here, so
long as we work to leading order in $1/N$ with $T$ scaled as $T \approx 1/N$.
We also have, from the Focker-Planck equations that the current
velocity is
\f
v^a_i (\lambda ) = {1 \over 2} (b^a_i + b^{*a}_i )
\label{lambdacurrent}
\ff
while the osmotic velocity is \f u^a_i (\lambda ) = {1 \over
2} (b^a_i - b^{*a}_i ) = \nu_\lambda {\delta \ln \rho_\lambda
\over \delta \lambda^a_i } \ff
>From these we can derive,
\f
 \lim_{\Delta t \rightarrow 0}   \int dQ
\rho_d (\lambda, t) \rho_Q (Q) {1\over 2} \left ( { (
\lambda^a_i (t + \Delta t ) - \lambda^a_i (t) )^2  \over \Delta
t^2 } \right ) =    \rho_d (\lambda, t)  [  b^a_i (\lambda , t)^2  +NC]
\ff and \f
 \lim_{\Delta t \rightarrow 0}   \int dQ
\rho_d (\lambda, t) \rho_Q (Q) {1\over 2}  \left ( { ( \lambda^a_i
(t  ) - \lambda^a_i  (t-\Delta t ) )^2  \over \Delta t^2 } \right
) =   \rho_d (\lambda, t) [ b^{*a}_i (\lambda , t)^2 +NC ]
\ff
where $C$ is the infinite constant we defined in eq. (\ref{C}).

To go
further we need to define the effective Hamilton-Jacobi function
for the eigenvalues. We define
\f
{S}_\lambda(\lambda ) = \int
(dQ) \rho_Q (Q) S(\lambda , Q)
\label{Slambda}
\ff
We now show that, to leading order in $1/N$,
\f
\mu v^a_i = { \delta S_\lambda (\lambda ) \over \delta
\lambda_a^i}.
\label{verygood}
\ff
Consider the probability
conservation law that follows from the statistical variational
principle that defines the dynamics of our matrix model, eq.
(\ref{mp}).
\f
\dot{\rho}(d,Q) = - { 1\over \mu} \left [ {\delta
\over \delta d_{ai}} ( \rho (d,Q) {\delta S(d,Q) \over \delta
d_{ai}} ) + {\delta \over \delta Q_{aij}} ( \rho (d,Q) {\delta
S(d,Q) \over \delta Q_{aij}} ) \right ]
\ff
But using (\ref{factorrho}) and (\ref{thermal}) we have that
\f
\dot{\rho}(d,Q) = \dot{\rho}_d (d) \rho_Q (Q).
\ff
We also have, by the same assumptions, since a thermal distribution is stationary
and has no current velocity,
\f
v^{aij}(Q) = {1 \over \mu} {\delta S(d,Q) \over \delta Q_{aij}} = O(1/N)
\ff
Thus, we have, integrating over the $Q$'s,
\f
\dot{\rho}_d (d) = -{1 \over \mu}{\delta \over \delta d_{ai}} ( \rho_d (d) {\delta \over \delta d_{ai}}
\int dQ \rho_Q (Q) S(d,Q) ) + O(1/N)      .
\ff
To leading order we can replace everywhere the dependence on $d_{ai}$ with dependence
on $\lambda_{ai}$, since the terms by which they differ are also higher order in
$1/N$.  Thus we have
\f
\dot{\rho}_d (\lambda ) = -{1 \over \mu}
{\delta \over \delta \lambda _{ai}} ( \rho_d (\lambda)
{\delta  S_\lambda (\lambda ) \over \delta \lambda_{ai}}    )
  + O(1/N)
\ff
But by (\ref{lambdacurrent}) we must have
\f
\dot{\rho}_d (\lambda ) = - {\delta \rho_d (\lambda ) v^{ai}(\lambda ) \over \delta \lambda_{ai}}
\ff
This establishes eq. (\ref{verygood}).
With this result we have the key relation that,
\f
{\mu \over 2} (b^2 + b^{*2} ) =
{\mu \over 2} (v^2 + u^2 ) = \left \{ {1\over 2\mu}  \left ( \delta S_\lambda
(\lambda ) \over \delta \lambda^a_i \right )^2 +  {\mu \nu^2_\lambda \over 2}
 \left (
\delta \ln \rho_\lambda (\lambda ) \over \delta \lambda^a_i \right
)^2 \right \}          .
\ff
We also define
\f
E_Q= \int dQ \rho_Q (Q) \left [
  {1\over 2\mu}({\delta S_Q(Q) \over \delta Q_{ij}^a} )^2
+ {\mu \omega^2 \over 2} Tr[Q,Q]^2
\right ]
\ff
and
\f
{\mu \over 2} \Omega_d^2 \sum_{aij} (\lambda^a_i
-\lambda^a_j )^2 = \int dQ \rho_Q (Q) U^{int} (\lambda, Q)
\ff
We can estimate that $E_Q = T N(N-1)/4 \approx N \mu \omega^2$ so this
is a divergent constant in the limit.
The
result is
\f
I[\rho_d, S,T] = \int dt \int  d\lambda [
\rho_d (\lambda, t) \dot{S}_\lambda -  H^{eff}(S_\lambda , \rho_d ,T)  ]
\ff
where,
the effective hamiltonian for the eigenvalues is
\f
 H^{eff}(S_\lambda , \rho_\lambda ,T)= \rho_d(\lambda ) \left [ E_Q^\prime   +
\left \{  \left (
{1 \over 2\mu} \delta S_\lambda (\lambda ) \over \delta \lambda^a_i \right )^2 +
{\mu \nu^2_\lambda \over 2}
 \left (
\delta \ln \rho_\lambda (\lambda ) \over \delta \lambda^a_i \right
)^2 \right \}
+ { \mu \Omega^2_d \over 2} \sum_{aij} (\lambda^a_I -\lambda^a_j )^2
 \right ]
\ff
where $E_Q^\prime = E_Q + N\mu C$ contains both infinite constants.
The resulting equations of motion are
\f
 E_Q^\prime + \dot{S}_\lambda +
 {1\over 2\mu }  \left (
\delta S_d (\lambda ) \over \delta \lambda^a_i \right )^2
+ { \mu \Omega^2_d \over 2} \sum_{aij} (\lambda^a_I -\lambda^a_j )^2 +
U^{quantum}   =  0
\ff
and the current conservation equation
\f
\dot{\rho}_\lambda = - {1\over \mu} \partial^{ai} \rho_\lambda ( \partial_{ai}
S_\lambda )
\ff
The so-called ``quantum potential" is given by
\f
U^{quantum} = \mu \nu^2_\lambda  \left \{ \left ( { \delta \ln
\rho_\lambda (\lambda ) \over \delta \lambda^a_i}  \right )^2 +
{1\over \rho_\lambda } \partial_{ai} ( \rho_\lambda \partial^{ai}
\ln \rho_\lambda ) \right  \} = - \mu \nu^2_\lambda {1\over \sqrt{
\rho_\lambda (\lambda )}} \nabla^2 \sqrt{ \rho_\lambda (\lambda )}
\ff
These we recognize as the real and imaginary parts of the
Schroedinger equation, when we write
\f
\Psi (\lambda, t) =
\sqrt{\rho_\lambda } e^{S_\lambda /\hbar}
\ff
with
\f
\hbar = \mu \nu_\lambda = \mu \omega {dt^{3/2}  \over 4(d-1)^{3/2}  }
\ff
So, finally,  we have in the limit $N \rightarrow \infty$,
\f
\imath \hbar {d \Psi(\lambda, t)  \over dt} = \left [ -{\hbar^2 \over 2\mu}
{\delta^2 \over \delta ( \lambda^a_i)^2} + { \mu \Omega^2_d \over 2} \sum_{aij}
(\lambda^a_I -\lambda^a_j )^2 + E_Q^\prime \right ] \Psi
(\lambda, t)
\ff
Finally, by repeating the argument of section 3.4 we can show that
the conserved energy of the original theory splits into two pieces,
\f
H=H^\Psi +E_Q^\prime
\ff
where
\f
H^\Psi = \int d\lambda \bar{\Psi}
\left ( - {\hbar^2 \over 2 \mu}  {\delta^2 \over \delta ( \lambda^a_i)^2}
+ { \mu \Omega^2_d \over 2} \sum_{aij} (\lambda^a_I -\lambda^a_j )^2
 \right ) \psi .
\ff Since $E_Q^\prime$ is an infinite constant the result is that
$H^\Psi$, which is the quantum mechanical energy, is conserved as
$N \rightarrow \infty$. We can then renormalize the wavefunctional
so that \f \Psi_r (\lambda ) = e^{iE_Q^\prime t /\hbar} \Psi
(\lambda ) \ff Finally, we note that as $\Omega^2_d \approx 1/N$
the eigenvalues become free in the limit $N \rightarrow \infty$.
Thus, when $N \rightarrow \infty$ the probabilities evolve
according to the free Schroedinger equation,
\f
\imath \hbar {d
\Psi_r(\lambda, t)  \over dt} = \left [ -{ \hbar^2 \over 2 \mu}
{\delta^2 \over \delta ( \lambda^a_i)^2} \right ] \Psi_r
(\lambda, t)
\ff

\section{Conclusions} What we have shown here may be summarized
by saying that matrix theory may not only give rise to string
theory, and hence gravity, it may also give rise to quantum
theory, in the sense that the quantum evolution of the eigenvalues
may appear, at large $N$, to be a consequence of the classical
statistical physics of the matrices. This means that we may be
able to solve the daunting conceptual problems of quantum theory
by means of a simple physical hypothesis: that the theory of
gravity and hence spacetime arises from a non-local background
independent theory in which geometry initially plays no role and
the physical degrees of freedom represent relational rather than
intrinsic properties. There remain many open questions. A short
list is: \begin{itemize} \item{} Is it possible to extend these
results to the actual supersymmetric models that arise in string
theory\footnote{This is in progress with Stephon
Alexander\cite{superthis}.}?

\item{}Is it possible that the existence of dualities that connect
certain quantum field theory observables to the classical limit of
string theory are related to the fact that the classical matrix
theory can in a certain limit reproduce a quantum theory?

\item{} How does Lorentz invariance and relativistic causality arise in the matrix models
which describe relativistic string and membrane theories, and how is this compatible
with the non-local dynamics of the off diagonal elements?

\item{}Can this be extended to truly background independent matrix models,such
as the cubic matrix models\cite{cubic} and the matrix models which have been
developed for spin foams\cite{foams}?

\item{}Are there any practical experimental predictions that follow from these
theories?

\end{itemize}
\section*{ACKNOWLEDGEMENTS}
I am very grateful first of all to Stephon Alexander for
discussions and collaboration on this idea and to Antony Valentini
for encouraging me to return to my earlier work on matrix models
as hidden variables theories. I also want to thanks Mark Davidson for
catching an error in an earlier draft of this paper. I am grateful to Chris Isham
for many conversations and for encouragement and support during my stay at Imperial College.
This work was supported by the
National Science Foundation and the Jesse Phillips Foundation, to
whom I am also grateful for encouragement to work on this problem.
Finally, I  am grateful to Lucian Hardy, Raymond Laflamme, Jaron Lanier,
Fotini Markopoulou, Rob Myers, Simon Saunders and Artem Starodubtrev for discussions on this model.

\end{document}